\documentstyle[preprint,aps,epsfig]{revtex}
\newcommand{\beq}{\begin{equation}}
\newcommand{\eeq}{\end{equation}}
\newcommand{\eq}[1]{Eq.(\ref{#1})}
\begin{document}
\draft
\preprint{DESY 98-150; hep-ph/9809277}
\tighten
\title {Hard Exclusive Electroproduction of Pseudoscalar Mesons
and QCD  Axial Anomaly }
\author {Michael I. Eides \thanks{E-mail address:  
eides@phys.psu.edu, eides@lnpi.spb.su}}
\address{ Department of Physics, Pennsylvania 
State University, 
University Park, PA 16802, USA,\\ 
and
Petersburg Nuclear Physics Institute,
Gatchina, St.Petersburg 188350, Russia}
\author{Leonid L. Frankfurt\thanks{E-mail address: frankfur@lev.tau.ac.il}}
\address{School of Physics and Astronomy, Raymond and Beverly Sackler
Faculty of Exact Sciences, \\Tel Aviv University, Ramat Aviv 69978, Israel
\and}
\author{\bigskip and \\Mark I. Strikman \thanks{E-mail address: 
strikman@phys.psu.edu}} 
\address{Department of Physics, Pennsylvania 
State University, University Park, PA 16802, USA,\\
and Deutsches Elektronen Synchrotron DESY, Germany\thanks{On leave of 
absence from PSU.}} 
\date{September, 1998}

\maketitle
\begin{abstract}

Hard exclusive electroproduction of $\pi^0,\eta,\eta'$ mesons by the 
longitudinal photons is considered.  With the help of the QCD factorization 
theorem we show that at sufficiently large virtualities $Q^2$ amplitudes of 
hard exclusive processes may be calculated in terms of the target parton 
distributions and the minimal Fock component of the pseudoscalar meson light 
cone wave function at zero interquark distances. This result justifies our 
calculation of the ratios of the $\pi^0 : \eta : \eta'$ yields for the 
electroproduction processes off nucleons and deuterons based on the low 
energy QCD chiral dynamics, and the QCD anomaly. Thus investigation of these 
ratios gives a new way to probe the physics of the spontaneous chiral 
symmetry breaking.  

\end{abstract} 
\pacs{PACS numbers: 13.60 Le} 

\newpage
\section{Introduction}

Recently  the QCD factorization theorem has been generalized to a large 
group of hard exclusive processes \cite{cfs}, see Eq. (\ref{process}) below.
On the experimental end, a group of hard exclusive high-energy processes was 
discovered at HERA and FNAL. Their properties agree with the hard QCD 
expectations but they are distinctively different from those observed in 
soft high energy processes, see \cite{fsdis} for recent summary and 
references. Discovery of these processes opened  a new frontier to probe 
the physics of the quark bound states and basic properties of QCD.  The aim 
of this paper is to demonstrate that exploration of the processes of hard 
exclusive electroproduction of pseudoscalar mesons off a hadron target gives 
a new way to probe the physics of the spontaneous chiral symmetry breaking.

Wealth of experimental data on the low-energy hadron dynamics was 
understood already before advent of QCD in terms of the current algebra 
and chiral Lagrangians (see, e.g., review in \cite{dgh}). After  emergence 
of QCD as the theory of strong interactions it turned out that predictions 
of the current algebra and the chiral Lagrangians could be more or less 
rigorously derived from the microscopic theory of strong interactions, 
if one assumes that the inherent chiral symmetry of QCD with massless 
quarks is spontaneously broken in the ground state. This bright picture was 
marred by the fact that naively spontaneous breaking of the chiral symmetry 
in QCD predicts a pattern of masses of the neutral pseudoscalar mesons 
$\pi,\eta,\eta'$ which contradicts the experimental data \cite{Weinberg}. 
This problem of the mass spectrum of the neutral pseudogoldstone bosons (and 
also of the large partial width of the decay $\eta\rightarrow 3\pi$)is 
referred to as the $U_1$ problem.  Resolution of the $U_1$ problem was 
triggered by the remark by t'Hooft \cite{hooft} that due to the 
Adler-Bell-Jackiw anomaly \cite{abj} the singlet axial current in QCD is 
non conserved and thus there could be a nontrivial contribution to the mass 
of the respective would be Goldstone boson.  Such additional contribution to 
the boson mass could be generated for example by the instanton 
contribution to the axial anomaly.  
A practical way to obtain a 
heavy $\eta'$ in QCD with large number of colors $N_c$ was suggested by 
Witten \cite{wit}. It is important to realize that the very existence of the 
axial anomaly is insufficient for resolution of the $U_1$ problem.  
Veneziano \cite{ven} has demonstrated that one has to assume existence of an 
 additional massless pole (so called ghost pole) in the correlators of the 
certain gauge-noninvariant gluon currents. It was checked a long time ago 
that the ghost pole hypothesis is self-consistent and provides a 
satisfactory explanation of the spectrum of the low-lying pseudoscalar 
mesons, of the large width of the decay $\eta\rightarrow 3\pi$, the widths 
of the decays $\eta,\eta'\rightarrow2\gamma$, and successfully describes 
also a number of other low-energy reactions \cite{de}.

We will show in  this paper that the ghost pole hypothesis 
leads to observable phenomena not only in the realm of the low-energy 
physics, but also in the hard exclusive electroproduction of pseudoscalar 
mesons. This means that one can probe the physics of the spontaneous chiral 
symmetry breaking in hard exclusive processes. We will obtain  a number of 
predictions for the ratios of certain cross sections for hard exclusive 
processes in terms of the parameters derived from the low-energy ghost-pole 
physics.

The key observation which leads to connection between the hard 
diffractive electroproduction and the parameters of the low-energy physics 
was made in \cite{cfs}, where the QCD factorization theorem was proven for 
the forward hard exclusive electroproduction of mesons 
by longitudinally polarized photons

\begin{equation} 
\gamma_L ^{*}(q) + p 
\to  M(q+\Delta ) + B'(p-\Delta ) \label{process} 
\end{equation} 
at large $Q^{2}$, with $t$ and $x=Q^{2}/2p\cdot q$ fixed.  
It asserts that the electroproduction amplitude has the form 

\begin{eqnarray} && \sum _{i} \int _{0}^{1}dz  \int 
d\xi f_{i/p}(\xi ,\xi -x;t,\mu ) \, H_{iM}(\xi /x,Q^{2},z,\mu ) \, 
\phi _{M}(z,Q^{2},\mu^2 ) \nonumber\\ && + \mbox{power-suppressed 
corrections}, 
\label{factorization} 
\end{eqnarray} 
where $f$'s are skewed 
(off(non)-diagonal(forward)) parton helicity densities within a nucleon, 
$\phi$ is the light-cone meson distribution amplitude \cite{lepbr}, 
$z$ is the light-cone fraction of the photon momentum carried by a quark 
(antiquark), and $H$ is a hard-scattering coefficient, which may be 
calculated in the form of the expansion in powers of the strong coupling 
constant $\alpha_{s}(Q)$. The scale $\mu^2 $ in Eq.(\ref{factorization}) as 
 usual for the factorization relations is $\sim Q^2$ hence there is a direct 
relation between meson wave functions probed in hard exclusive two-body 
processes and in form factor physics.

Note that for sufficiently small $x \le 0.1$ and large 
$Q^2$ the skewed distributions can be expressed through the diagonal 
distributions at large $x$ and  $Q^2 \sim $ few GeV$^2$ 
\cite{Freund-Frankfurt-Gusey-Strikman,martin}. Relation between skewed and 
usual parton densities at moderately small $x$  has been discussed in 
\cite{radyushkin}.  At the same time  for $x \ge 0.1$ and  $Q^2 \sim $ few 
GeV$^2$ calculations of the  skewed distributions within the chiral model 
were recently performed in \cite{bochum}.

Moreover these skewed  parton densities  really will not enter into our 
final expressions for the ratios of cross sections since they are  almost 
completely canceled out  in these ratios. This cancellation is especially 
precise in the case of the target with vanishing isospin, namely for the 
deuteron target. We neglect here the contribution of the difference between 
the  $s$ and $\bar s$  distributions within a nucleon target. Thus for large 
$Q^2$ only the integral 

\begin{equation}
\int \frac{dz}{z(1-z)} \phi_M(z,Q^2,\mu^2=Q^2).
\label{ampl}
\end{equation}
of the light-cone meson distribution amplitude
$\phi_M(z,Q^2,\mu^2)$ enters in the expressions for the ratios 
of cross sections of electroproduction of pseudoscalar mesons.
 
At the same time it is well known that the meson distribution amplitude
is completely determined by the conformal invariance of QCD and the quantum 
numbers of the meson $M$ and can be decomposed in a series over the 
Gegenbauer polynomials $C^\frac{3}{2}_n$ \cite{lepbr,efrrad,chzh}. In the 
limit $Q^2\rightarrow\infty$ only the leading term with $n=0$ survives and 
\begin{equation}
\phi_M(z,Q^2,\mu^2=Q^2)=c_Mz(1-z).
\end{equation}
As a result in the case of pseudoscalar 
mesons integration over $z$ in the expression for the amplitude becomes 
trivial and the amplitude in \eq{ampl} is simply proportional to $c_M$. 
Hence, all  nontrivial dynamical information about the amplitude is 
contained in this case in the normalization factor of the light-cone wave 
function which is equal to the matrix element of the local axial current 
\cite{lepbr,efrrad,chzh}. 
In this respect calculation of ratios of hard diffractive pseudoscalar meson 
production cross sections is theoretically even more transparent than 
calculation of absolute cross sections at moderate $Q^2$, when respective 
integrals of the meson wave function are more complicated and do not reduce 
to  matrix elements of a local operator \cite{cfs}\footnote{
It is worth noting that experimental investigation of the ratios of
the yields of different mesons has another important advantage.
Indeed, it was demonstrated in \cite{FKS} that up to rather
large $Q^2$ the absolute
cross sections are suppressed significantly due to the higher twist effects
originating from comparable transverse size of the longitudinal
photon and meson wave functions. However the overall transverse 
size was found to be sufficiently small $\le 0.4 fm$ at $Q^2 \ge$ 5 GeV$^2$.
Due to color transparency this leads to  a strong suppression of the
final state
interaction of the  $q \bar q$ pair which will form the meson 
and the residual baryon system. For $W \le 20 GeV$ this cross section is 
of the order of few $mb$. Besides, the expansion of the $q \bar q $ system 
to a normal hadron size in the nucleon rest frame
takes a distance $l_{coh} \sim p_M/\Delta m_M^2$ where $\Delta m_M^2$
is characteristic light-cone energy denominator for a meson $M$ 
which is $\le$ 1 GeV$^2$,
cf. review \cite{fms}.
 One can easily see 
that condition $\l_{coh} \gg r_N$
is  satisfied for $x \ge 0.2$ already for $Q^2 \ge$ 5 GeV$^2$.
Hence, it seems likely 
that {\it a precocious factorization into three blocks} -
overlap integral, hard blob, skewed density - could be valid
already at relatively low Q$^2$ leading to {\it a precocious
scaling of the ratios of the cross sections} as a function $Q^2$.
At the same time predictions for the  absolute cross sections
are valid only for  $Q^2 \geq 10 GeV^2$.
In the case of the ratios of the yields  of pseudoscalar and vector mesons, 
like the $\pi^0/\rho^0$ ratio calculated in \cite{mpw,vgg} at large $Q^2$ 
this argument may not work as well due to possible different
transverse sizes of $q \bar q$ configurations for pseudoscalar and vector 
mesons.}.

Let us emphasize once again that the same matrix element of the axial 
current determines both the amplitudes of the low-energy interaction of the 
pseudoscalar mesons and the amplitudes of the hard diffractive 
electroproduction of these mesons. Thus the hard diffractive processes can 
be used to investigate intricate properties of the physics of spontaneously 
broken chiral symmetry in QCD.

\section{ Spin dependence  of the amplitudes for pseudoscalar meson
electroproduction in QCD}

Quark-antiquark pairs in hard exclusive processes initiated by 
longitudinally polarized photons are produced at  small relative
transverse distances $\propto 1/Q$. Thus such processes probe interaction 
of a small size $q \bar q$ pair with a hadron target. Localization of the 
color charge in a small volume justifies applicability of the perturbative 
QCD for the description of such processes. Thus to describe hard exclusive 
electroproduction of pseudoscalar mesons it is necessary to calculate the
handbag diagram. Convenient technology of calculation is to evaluate 
the imaginary parts of the amplitudes and then to reconstruct the real part 
using the dispersion representation of the amplitude over energy. 
Simplification arises since the amplitude corresponding to the negative 
charge parity in the crossed channel has no subtraction constants and is 
unambiguously determined by the dispersion relations. The gluons which are 
produced in the intermediate state are predominantly transversely polarized 
relative to the reaction axis defined by the photon and the target momenta.
It is convenient to use the Fierz transformation to factorize the spin 
structure of the amplitudes $\sum_{\mu=1,2}\gamma_{\mu} 
\otimes\gamma_{\mu}=2 \gamma_5\otimes\gamma_5 
+\gamma_{\mu_{\|}}\gamma_5\otimes\gamma_{\mu_{\|}}\gamma_5 +\ldots$. 
We omit terms which do not give  contribution to the box diagram.  In the  
case of forward scattering the first term gives vanishing contribution as a 
result of chirality conservation. For nonforward scattering when $t$ is not 
large the second term dominates also since the contribution of the first 
term to the amplitude of electroproduction of a pseudoscalar meson is 
suppressed by an additional factor $x$. Thus we have proved that only the 
spin flip amplitude gives nonvanishing contribution (see also 
\cite{cfs,mpw}). Factorization of the spin structure of the handbag diagram 
from that in the parton distributions within a target means that the handbag 
diagram is effectively averaged over the quark helicities. Due to the 
dominance of the $\gamma_{\mu_{\|}}\gamma_5$ term only the component of the 
meson wave function which is given by  the matrix element of the axial  
current survives in the trace of the handbag diagram after this averaging.

\section{Naive $SU(3)$ Symmetric Relative Production Rates}

As we have emphasized above the interquark  distances in the pseudoscalar 
mesons probed in hard processes are small. Hence, the decomposition of the 
hadron state vector in terms of the quark and gluon Fock space states can 
be mathematically defined and justified within the framework of the  QCD 
factorization theorem. In the case of hard diffractive electroproduction of 
a meson by longitudinally polarized photons only the quark-antiquark 
component  of the general wave function contributes\cite{bfgms}. Other 
partons of the light cone wave function of a hadron are included into the 
structure functions of a target.  The amplitude for the exclusive neutral 
pseudoscalar production off the hadron target in terms of the light cone 
wave functions were obtained in \cite{cfs}.  Weights of quarks of different 
flavors in the box diagram are determined by the electromagnetic current 
operator 
\beq j_\mu=\frac{2}{3}\bar u\gamma_\mu u-\frac{1}{3}\bar 
d\gamma_\mu d -\frac{1}{3}\bar s\gamma_\mu s.  
\eeq 
On the other hand the strength of the pseudoscalar-quark-antiquark vertex is 
determined, as we have explained above, by the matrix elements 

\beq  \label{axial}
<0|\bar\psi_f\gamma^\mu\gamma_5\psi_f|M> =if_f p^\mu,
\eeq
where $\psi_f$ is a quark field of flavor $f$. 

Hence, the helicity distribution functions of the quarks of flavor 
$f$ enter the production amplitudes with the weights determined by the
effective axial current
\beq
j_\mu^5=i(\frac{2}{3}\bar u\gamma_\mu\gamma_5 u
-\frac{1}{3}\bar d\gamma_\mu\gamma_5 d
-\frac{1}{3}\bar s\gamma_\mu\gamma_5 s),
\eeq
or in other words the weights are proportional to the matrix elements in 
\eq{axial} multiplied by the respective quark electric charges.  

Let us consider first the hard exclusive neutral pseudoscalar production in 
the limit of exact $SU(3)$ symmetry  neglecting the axial anomaly. In 
this case matrix elements of the $SU(3)$ axial currents 
\beq      \label{pion}
<0|
\frac{\bar u\gamma^\mu\gamma_5u-
\bar d\gamma^\mu\gamma_5d}{\sqrt2}
|\pi^0> =if_\pi p^\mu,
\eeq
\[
<0|\frac{\bar d\gamma^\mu\gamma_5d+\bar u\gamma^\mu\gamma_5 u
-2\bar s\gamma^\mu\gamma_5 s}{\sqrt{6}}
|\eta> =if_\eta p^\mu,
\]
\[ 
<0|\frac{\bar d\gamma^\mu\gamma_5 d+\bar u\gamma^\mu\gamma_5 u
+\bar s\gamma^\mu\gamma_5 s}{\sqrt{3}}
|\eta'> =if_{\eta'} p^\mu,
\]
contain one and the same constant $f_\pi=f_\eta=f_{\eta'}\approx 132$ MeV 
on the right hand side. The matrix elements in \eq{axial} can be easily 
obtained from the $SU(3)$ relationships

\beq      \label{pionmatr}
<0|\frac{\bar u\gamma^\mu\gamma_5u}{\sqrt2}
|\pi^0> =-<0|\frac{\bar d\gamma^\mu\gamma_5d}{\sqrt2}
|\pi^0> =i\frac{f_\pi}{2} p^\mu,
\eeq
\beq      \label{etamatr}
<0|\frac{\bar d\gamma^\mu\gamma_5d}{\sqrt{6}}|\eta>
=<0|\frac{\bar u\gamma^\mu\gamma_5 u}{\sqrt{6}}|\eta>
=-<0|\frac{\bar s\gamma^\mu\gamma_5 s}{2\sqrt{6}} |\eta> =i\frac{f_\eta}{6}
p^\mu,
\eeq
\beq      \label{etaprimematr}
<0|\frac{\bar d\gamma^\mu\gamma_5
d}{\sqrt{3}}|\eta'>=<0|\frac{\bar u\gamma^\mu\gamma_5 u}{\sqrt{3}}|\eta'>
=<0|\frac{\bar s\gamma^\mu\gamma_5 s}{\sqrt{3}} |\eta'>
=i\frac{f_{\eta'}}{3} p^\mu.
\eeq                                                         
In this approximation the amplitudes of hard exclusive diffractive 
electroproduction have the form (we omit the common factor $f_\pi$ 
since we are interested only in the ratios of the amplitudes)
\beq   \label{prodampl}
<0|j_\mu^5|\pi^0>\sim\frac{1}{\sqrt{2}}(\frac{2}{3}\Delta
u+\frac{1}{3}\Delta d),
\eeq
\[
<0|j_\mu^5|\eta>\sim\frac{1}{\sqrt{6}}(\frac{2}{3}\Delta
u-\frac{1}{3}\Delta d+\frac{2}{3}\Delta s),
\]
\[
<0|j_\mu^5|\eta'>\sim\frac{1}{\sqrt{3}}(\frac{2}{3}\Delta
u-\frac{1}{3}\Delta d-\frac{1}{3}\Delta s), 
\]
where $\Delta u$, $\Delta d$ and $\Delta s$ are the helicity distribution 
functions of the quarks in the target.  Hence in the SU(3) limit we obtain:
\beq   \label{ratio1}
\pi^0:\eta:\eta'=
\frac{1}{2}\left(\frac{2}{3}\Delta
u+\frac{1}{3}\Delta d\right)^2:
\frac{1}{6}\left(\frac{2}{3}\Delta
u-\frac{1}{3}\Delta d+\frac{2}{3}\Delta s\right)^2:
\frac{1}{3}\left(\frac{2}{3}\Delta
u-\frac{1}{3}\Delta d-\frac{1}{3}\Delta s\right)^2, 
\eeq
Assuming that for the proton $\Delta 
s=0$ and $\Delta d\approx -\Delta u$, we immediately obtain the ratios of 
production cross sections
\beq         \label{ratio}
\pi^0:\eta:\eta'=1:3:6,
\eeq
which were first obtained in \cite{cfs}. Above ratios differ from those 
obtained in \cite{mpw} within the same approximations. A small uncertainty 
in the predictions due to unknown value of $\Delta s$ can be further 
suppressed by selecting $x \ge 0.2$ where the strange sea is surely 
negligible.

\section{Relative Production Rates with Account of the
QCD Axial Anomaly and $SU(3)$ Breaking}

Due to the presence of the QCD axial anomaly in the neutral channel, and 
due to an explicit flavor $SU(3)$ breaking by the current quark masses, the 
matrix elements of the axial currents in QCD have a more involved structure 
than in the $SU(3)$ symmetric case considered above. These two effects in 
the matrix elements were taken into account in \cite{de}, and we will use 
these results in the problem of hard exclusive diffractive production of the 
pseudoscalar mesons.

The relationships in \eq{pionmatr} remain valid since we still assume
$SU(2)$ symmetry and there is no anomaly in the 
isovector channels. In order to
obtain the axial current matrix elements for $\eta$ and $\eta'$ we introduce 
auxiliary currents
\beq
j^1_{\mu5}=\frac{\bar u\gamma_\mu\gamma_5 u+\bar d\gamma_\mu\gamma_5
d}{\sqrt{2}},
\eeq
\beq
j^2_{\mu5}=\bar s\gamma_\mu\gamma_5 s.
\eeq
Conservation of these currents is broken by the QCD axial anomaly
\beq     \label{anomaly}
\partial_\mu j^1_{\mu5} =2i\frac{m_u\bar u \gamma_5 u+m_d\bar d \gamma_5
d}{\sqrt{2}}+2\sqrt{2}Q\equiv P_1+2\sqrt{2}Q,
\eeq
\[
\partial_\mu j^2_{\mu5} =2im_s\bar s \gamma_5 s
+2Q\equiv P_2+2Q,
\]
where $Q=\frac{\alpha_s}{8\pi}G\tilde G$.

We need to obtain analogies of the relationships in \eq{etamatr} and 
\eq{etaprimematr} in the real QCD. As a technical tool we 
define new constants $f^1_{\eta(\eta')}$ and $f^2_{\eta(\eta')}$ (note 
that these constants $f^i_{\eta(\eta')}$ do not coincide with any constants 
in \cite{de})

\beq       \label{anomconst}
<0|j^1_{\mu5}|\eta> =if^1_\eta p^\mu,
\eeq
\beq      
<0|j^1_{\mu5}|\eta'> =if^1_{\eta'} p^\mu,
\eeq
\beq      
<0|j^2_{\mu5}|\eta> =if^2_{\eta} p^\mu,
\eeq
\beq      
<0|j^2_{\mu5}|\eta'> =if^2_{\eta'} p^\mu.
\eeq

In the end of the day these constants will determine the production
amplitudes.

Due to the axial anomaly in \eq{anomaly} we have

\beq  \label{f1eta}
f^1_\eta m_\eta^2=<0|P_1+2\sqrt{2}Q|\eta>,
\eeq
\beq  \label{f1eta1}
f^1_{\eta'} m_{\eta'}^2=<0|P_1+2\sqrt{2}Q|\eta'>,
\eeq
\beq  \label{f2eta}
f^2_\eta m_{\eta}^2=<0|P_2+2Q|\eta>,
\eeq
\beq  \label{f2eta1}
f^2_{\eta'} m_{\eta'}^2=<0|P_2+2Q|\eta'>.
\eeq
Expressions on the right hand side of these equations were calculated in 
\cite{de} (note that $f_1$ and $f_2$ in the relationships below are as they 
are defined in \cite{de} and have nothing to do with the numerous $f$'s 
defined above)

\beq
<0|Q|\eta>=\sqrt{\frac{\lambda^4(m_2^2-m_\eta^2)(m_\eta^2-m_1^2)}
{m_{\eta'}^2-m_\eta^2}},
\eeq
\[
<0|Q|\eta'>=\sqrt{\frac{\lambda^4(m_{\eta'}^2-m_2^2)(m_{\eta'}^2-m_1^2)}
{m_{\eta'}^2-m_\eta^2}},
\]
\[
<0|P_1|\eta>=\sqrt{\frac{f_1^2m_1^4(m_2^2+\mu_2^2-m_\eta^2)}
{m_{\eta'}^2-m_\eta^2}},
\]
\[
<0|P_1|\eta'>=\sqrt{\frac{f_1^2m_1^4(m_{\eta'}^2-m_2^2-\mu_2^2)}
{m_{\eta'}^2-m_\eta^2}},
\]
\[
<0|P_2|\eta>=-\sqrt{\frac{f_2^2m_2^4(m_1^2+\mu_1^2-m_\eta^2)}
{m_{\eta'}^2-m_\eta^2}},
\]
\[
<0|P_2|\eta'>=\sqrt{\frac{f_2^2m_2^4(m_{\eta'}^2-m_1^2-\mu_1^2)}
{m_{\eta'}^2-m_\eta^2}}.
\]

Now we are in a position to calculate unknown constants
$f^{(i)}_{\eta(\eta')}$. We use parameters from  \cite{de}

\beq
f_1=f_\pi=0.132 GeV,\qquad f_2=f_\pi+2(f_K-f_\pi)=0.178 GeV,\qquad
\frac{f_2}{f_\pi}=1.35
\eeq
\[
\mu_1^2=0.57 GeV^2,\qquad \mu_2^2=0.16 GeV^2,
\qquad \lambda^4=\frac{f_\pi^2\mu_1^2}{8}=(0.188 GeV)^4
\]
\[
m_\eta^2=0.307
GeV^2 \;(0.301),\qquad m_{\eta'}^2=0.912 Gev^2\; (0.917),
\]
\[
m_1^2\approx m_\pi^2\approx 0.02 GeV^2,\qquad m_2^2\approx
m_{K^+}^2+-m_\pi^2\approx 0.47 GeV^2,
\]
where in the brackets we have written experimental values of the masses.

Note that if one ignores the anomalous contribution and
the $SU(3)$ breaking for the constants $f_i$ then our new formula 
immediately reproduce the naive $SU(3)$ symmetric results  
considered in \cite{cfs} and reproduced above.

Substituting the numbers above in \eq{f1eta}-\eq{f2eta1} 
we obtain the numerical values for the constants $f^i_{\eta(\eta')}$ 

\beq
f^{(1)}_\eta=0.097,\qquad f^{(1)}_{\eta'}=0.090,
\eeq
\[
f^{(2)}_{\eta}=-0.122,\qquad f^{(2)}_{\eta'}=0.129.
\]

Using \eq{anomconst} we obtain

\beq      
<0|j^1_{\mu5}|\eta> =0.097ip^\mu,
\eeq
\beq      
<0|j^1_{\mu5}|\eta'> =0.090i p^\mu,
\eeq
\beq      
<0|j^2_{\mu5}|\eta> =-0.122i p^\mu,
\eeq
\beq      
<0|j^2_{\mu5}|\eta'> =0.129i p^\mu,
\eeq

or

\beq      
<0|\frac{\bar u\gamma_\mu\gamma_5 u}{\sqrt{2}}|\eta>
=<0|\frac{\bar d\gamma_\mu\gamma_5
d}{\sqrt{2}}|\eta> =\frac{0.097}{2}i p^\mu
=0.366if_\pi p^\mu, \qquad (0.289),
\eeq
\beq      
<0|\frac{\bar u\gamma_\mu\gamma_5 u}{\sqrt{2}}|\eta'>
=<0|\frac{\bar d\gamma_\mu\gamma_5
d}{\sqrt{2}}|\eta'> =\frac{0.090}{2}i p^\mu
=0.343if_\pi p^\mu, \qquad (0.408),
\eeq
\beq      
<0|\bar s\gamma_\mu\gamma_5 s|\eta> =-0.122i p^\mu
=-0.927if_\pi p^\mu, \qquad (-0.816),
\eeq
\beq      
<0|\bar s\gamma_\mu\gamma_5 s|\eta'> =0.129i p^\mu
=0.980if_\pi p^\mu, \qquad (0.577),
\eeq
where the naive values of the coefficients from \eq{pionmatr} are written in 
the parentheses.

Now we can easily repeat calculation of the production amplitudes and cross
sections made above with account of the anomaly and the $SU(3)$
breaking. We obtain instead of \eq{prodampl} (note that nothing changes for
$\pi^0$-meson)
\beq
<0|j_\mu^5|\pi^0>\sim\frac{1}{\sqrt{2}}(\frac{2}{3}\Delta
u+\frac{1}{3}\Delta d),
\eeq
\[
<0|j_\mu^5|\eta>\sim \frac{2\sqrt{2}}{3}0.366\Delta
u-\frac{\sqrt{2}}{3}0.366\Delta d+\frac{1}{3}0.927\Delta s
\]
\[
=\frac{1}{\sqrt{6}}(\frac{2}{3}\cdot1.27\Delta
u-\frac{1}{3}\cdot1.27\Delta d+\frac{2}{3}\cdot1.14\Delta s),
\]
\[
<0|j_\mu^5|\eta'>\sim\frac{2\sqrt{2}}{3}0.343\Delta
u-\frac{\sqrt{2}}{3}0.343\Delta d-\frac{1}{3}0.980\Delta s
\]
\[
=\frac{1}{\sqrt{3}}(\frac{2}{3}\cdot0.840\Delta
u-\frac{1}{3}\cdot0.840\Delta d-\frac{1}{3}\cdot1.697\Delta s). 
\]
Then  for the cross section ratios we have (compare with the $SU(3)$ 
symmetric result in \eq{ratio1})
\beq  \label{36}
\pi^0:\eta:\eta'=
\frac{1}{2}\left(\frac{2}{3}\Delta
u+\frac{1}{3}\Delta d\right)^2:
\frac{1}{6}\left(\frac{2}{3}\cdot1.27\Delta
u-\frac{1}{3}\cdot1.27\Delta d+\frac{2}{3}\cdot1.14\Delta s\right)^2
\eeq
\[
:\frac{1}{3}\left(\frac{2}{3}\cdot0.840\Delta
u-\frac{1}{3}\cdot0.840\Delta d-\frac{1}{3}\cdot1.697\Delta s\right)^2.
\]

To analyze the consequences of \eq{36} we will use the standard 
assumption that the quark helicity distributions satisfy the relation 
$\Delta s=0$, see e.g. \cite{Florian}.

First we observe that the  $\eta$ production rate is enhanced by a factor of 
1.61 and the $\eta'$ production is suppressed by a factor of 0.71 in 
comparison with the naive $SU(3)$ limit, leading to 

\beq        
\label{crospred} \eta:\eta'=1:0.87 
\eeq 
to be compared with the naive result 1:2 in \eq{ratio}.  Qualitatively 
this is quite natural since it means enhancement of the $\eta$ coupling with 
the $u$ and $d$ quarks, and suppression of the $\eta'$ coupling with the 
same quarks. This may be interpreted in the sense that due to mixing with 
the QCD ghost pole $\eta'$-meson contents is enriched by additional glue.  

In the case of the coherent  scattering off the deuteron
large cancellations occur since the amplitudes for scattering off proton and 
neutron have opposite sign due to $\Delta u \approx -\Delta d$.  This 
indicates that coherent diffraction off the deuteron will be strongly 
suppressed as compared to the break up channel.  There is an additional 
suppression due to a spin flip in the deuteron vertex.  Naive $SU(3)$ 
symmetric result for the cross sections ratios for the coherent scattering 
off the deuteron is given by the relationship $\pi:\eta:\eta'=27:1:2$, while 
the calculation with account of the chiral anomaly and $SU(3)$ breaking 
leads to the result \beq \pi:\eta:\eta'=27:1.6:1.4, \eeq which is markedly 
different from the case of scattering off the proton as well as off the
 neutron (which can be studied in the deuteron break up reaction).

For numerical studies of the implications of \eq{36} for production of 
$\eta$ and $\pi^0$ off nucleons we use the results of the recent global 
analysis of inclusive and semi-inclusive polarization data \cite{Florian}. 
For the first estimate we neglect effects of skewedness of relevant parton 
densities. For certainty we fix $Q^2= 20 GeV^2$ and take the fit 1 of the 
set of fits presented in \cite{Florian}. We checked that our results depend 
very weakly on $Q^2$ scale of the parton densities. This is due to a rather 
small scaling violation in the valence quark channels. Also we found that 
different fits 
of \cite{Florian}
 lead to rather small ($\le 10\%$) changes of the 
considered ratios.

In Fig.1 we present the ratio of the cross section of production of $\eta $ 
and $\pi^0$ off the proton and neutron targets.  In Fig.2 we compare yields 
of $\pi^0$ and $\eta$ off the proton and neutron targets. One can see that 
the current data on $\Delta u_V$, $\Delta d_V$ lead to a rich pattern of 
$x$, and target dependencies which would be worth studying experimentally.  
We also observe that a simplifying approximation $\Delta u_V \approx - 
\Delta d_V$ is inadequate for the quantitative predictions.

A set of axial current matrix elements $<0|\bar 
q\gamma_{\mu}\gamma_5q|\eta>$, $<0|\bar q\gamma_{\mu}\gamma_5q|\eta>$ was 
obtained in a recent phenomenological analysis \cite{kroll} of the various 
low energy data on production and decays of $\eta, \eta'$ mesons. The 
numbers derived in \cite{kroll} differ strongly from the SU(3) expectations, 
and are pretty close to the ones obtained in \cite{de} from the analysis of 
the anomalous Ward identities in QCD and used in the present work. The 
largest difference ($\sim 13\%$) between the two sets of matrix elements 
occurs for the $<0|\bar u\gamma_{\mu}\gamma_5u|\eta>$ matrix element. In 
case of the cross section ratios considered above the phenomenological 
parameters from \cite{kroll} would decrease the value of the  $\eta'/\pi^0$ 
ratio by a factor of .98 and increase the value of the $\eta/\pi^0$ ratio by 
a factor $\sim 1.27$.  

Similar matrix elements are relevant  also for the description of the 
processes $\gamma\gamma^*\to \pi^0,\eta,\eta'$ at large $Q^2$\cite{BL}.  In 
the case of the $\pi^0$ production the QCD analysis of \cite{radyushkin2} 
has provided a good description of the data giving a new evidence for 
closeness of the $q\bar q$ component of the pion wave function to the 
asymptotic shape.  Analysis of the $\eta,\eta'$ production performed 
recently in \cite{kroll} shows that the large $Q^2$ data agree with the 
results of calculations based on the values of the phenomenological matrix 
elements obtained in \cite{kroll}. The matrix elements from \cite{de} lead 
to the same magnitude of the $\eta'$ transition form factor at large $Q^2$ 
and to about $\sim 20\%$ lower magnitude of the $\eta$ transition form 
factor.  The accuracy of the data is not sufficient yet to discriminate 
between the two sets of the matrix elements.

\section{Discussion of Results}

We have demonstrated above that the hard exclusive electroproduction of the 
pseudoscalar mesons $M$ off a nucleon  $\gamma +N\rightarrow M+N$ can be 
described in the leading twist with the help of the perturbative quark 
handbag diagram. However, this production process may be  also mediated by 
the odderon exchange \cite{odderon}. Distinctive feature of the odderon 
exchange is that the cross section only slowly depends on the photon energy 
but rapidly decreases as $\alpha_s^3(Q^2)/{Q^{10}}$ with increase of the 
photon virtuality $Q^2$. On the other hand contribution to the cross section 
which is due to the exchange of $q\bar q$ pair discussed above behaves 
itself as $\alpha_s^2 x^n/(Q^6)$, i.e. this contribution decreases with 
energy.  Here, $n$ is equal 2 if the perturbative $q\bar q$ pair exchange 
describes amplitude at large intervals in rapidity.  If the $x$ dependence 
of amplitude  due to the  q$\bar q$ pair exchange is estimated  assuming 
that the $x$ dependence of  the nonperturbative support function in the QCD 
evolution equation is given by the exchange by the vector meson Regge 
trajectory, then $n \approx 1$.

The ratio of the respective cross sections has 
the form  $c\alpha_s(Q^2)/Q^4 x^n$, where $c$ is a model dependent numerical 
factor. Odderon exchange is a sum of the hard and soft QCD 
contributions.  Even signs of these contributions cannot be derived from 
the general principles. Moreover, hard contribution to the odderon exchange 
 cannot be expressed through the parton distributions of a proton 
since the soft (nonperturbative) end-point contribution in the integration 
over the fraction of the photon momentum carried by the quark lines is not 
suppressed in the case of the odderon exchange.

One option to suppress uncontrollable and therefore dangerous odderon 
contribution is to investigate the ratios at moderate $x$ where the odderon 
exchange is suppressed.  Another option is to consider the  processes where 
the odderon contribution is forbidden by the quantum numbers. A reaction 
with $\Delta^+$ production on the proton $\gamma_L+p\rightarrow M+\Delta^+$ 
is a good example. In this case the odderon exchange is suppressed by the 
isospin conservation.  Another advantage of this reaction is that in this 
case the poorly understood distribution of strange quarks within a proton 
gives no contribution. Hence, the amplitude of this production process is 
expressed exclusively in terms of the valence quark distributions. The 
relation between the electroproduction cross sections of $\pi, \eta,\eta'$ 
in this reaction may be easily calculated in the same way as it was done 
above and one may prove that it exactly coincides with the one obtained in 
\eq{crospred}.

\medskip

In conclusion, we have demonstrated that the ratios of cross sections 
of hard exclusive electroproduction of the neutral pseudoscalar mesons may 
be calculated in terms of the low-energy physics. This low-energy physics 
crucially depends on the QCD axial anomaly and our results mean that one may 
explore the physics of spontaneous chiral symmetry breaking in the 
high-energy processes.

\acknowledgements

We are deeply grateful to A. Radyushkin for useful discussion, to T. 
Feldmann for drawing our attention to the papers \cite{kroll},  and 
to S. Bass and  M. Moinester for discovering a misprint in the 
first draft of the paper.

This work was supported by the U.S. Department of Energy Under Contract 
No. DE-FG02-93ER-40771, and by the Israel Academy of Science  under contract
N 19-971.

\newpage

\begin{figure}
\centerline{\epsfig{file=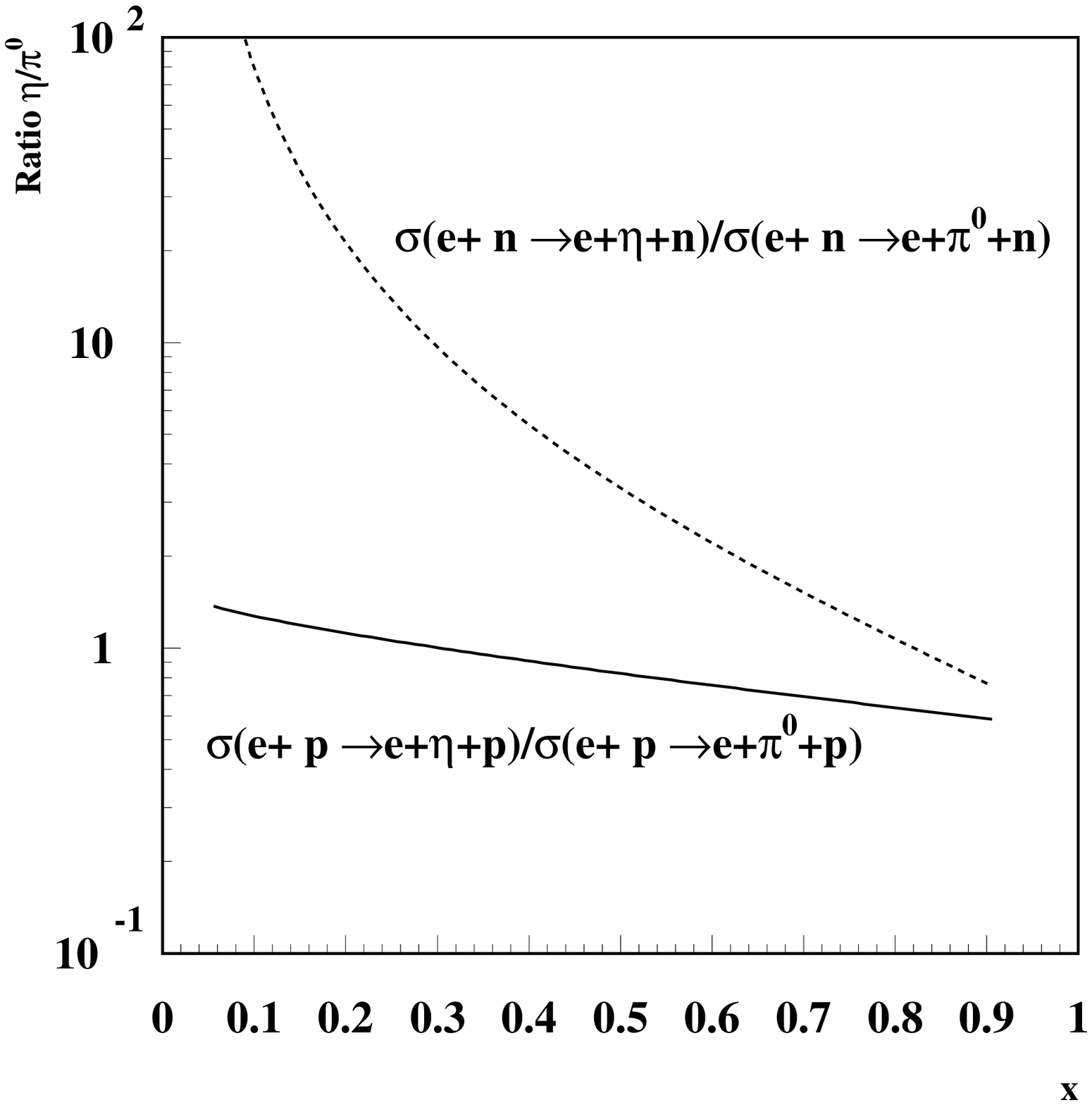,width=15.0cm,height=15.0cm}}
\caption {The $x$ dependence of the ratio 
$\sigma(\gamma_L+N \to \eta +N)/\sigma(\gamma_L+N \to \pi^0 +N)$
 for the proton and neutron targets calculated using
\eq{36} and parameterization of $\Delta u_V$ and
$\Delta d_V$ from \protect\cite{Florian}.}
\end{figure}
\newpage
\begin{figure}
\centerline{
\epsfig{file=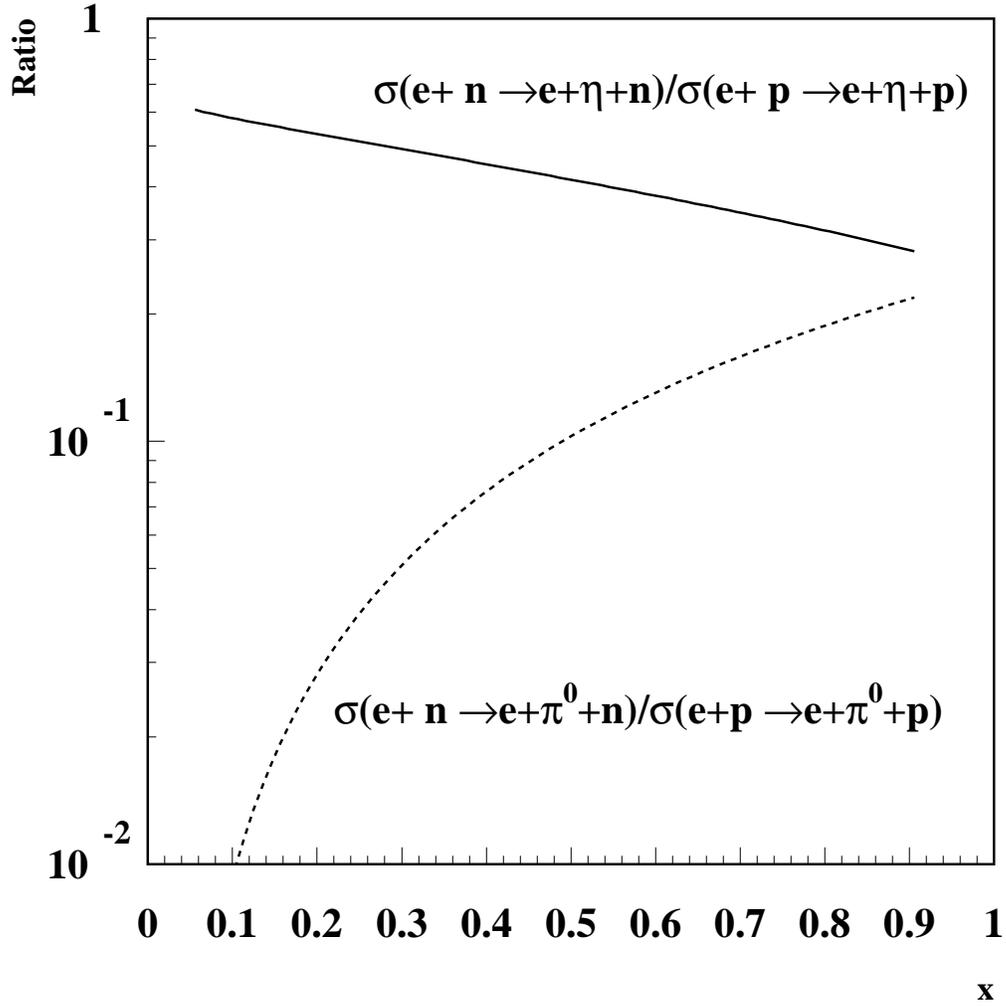,width=15.0cm,height=15.0cm}}
\caption {The $x$ dependence of the ratio 
$\sigma(\gamma_L+n \to M +n)/\sigma(\gamma_L+p \to \pi^0 +p)$
for production of $\eta$, and $\pi^0$ mesons calculated using 
\eq{36} and parameterization of $\Delta u_V$ and
$\Delta d_V$ from \protect\cite{Florian}.}
\end{figure}
\end{document}